\documentclass[preprint,12pt]{elsarticle}

\usepackage{amssymb}

\usepackage{lineno}

\usepackage{epsfig}
\usepackage{subfigure}
\usepackage{color}
\usepackage{bm}
\usepackage{latexsym}

\usepackage{epsfig}

\usepackage{graphicx}
\usepackage{epstopdf}

\usepackage{mathtools}
\usepackage{amsfonts}

\usepackage{float}

\journal{Chaos Soliton Fract.}

\begin{document}

\newcommand{\bra}{\left\langle}
\newcommand{\ket}{\right\rangle}
\newcommand{\tbox}[1]{\mbox{\tiny #1}}

\newcommand{\hide}[1]{}
\newcommand{\half}{\mbox{\small $\frac{1}{2}$}}
\newcommand{\sinc}{\mbox{sinc}}
\newcommand{\const}{\mbox{const}}
\newcommand{\trc}{\mbox{trace}}
\newcommand{\intt}{\int\!\!\!\!\int }
\newcommand{\ointt}{\int\!\!\!\!\int\!\!\!\!\!\circ\ }
\newcommand{\eexp}{\mbox{e}^}
\newcommand{\EPS} {\mbox{\LARGE $\epsilon$}}
\newcommand{\ar}{\mathsf r}
\newcommand{\im}{\mbox{Im}}
\newcommand{\re}{\mbox{Re}}
\newcommand{\bmsf}[1]{\bm{\mathsf{#1}}}
\newcommand{\mpg}[2][1.0\hsize]{\begin{minipage}[b]{#1}{#2}\end{minipage}}

\begin{frontmatter}



\title{Leaking from the phase space of the \\ 
Riemann-Liouville fractional standard map}


\author{J. A. M\'endez-Berm\'udez}
\address{Instituto de F\'isica, Benem\'erita Universidad Aut\'onoma de Puebla, 
Apartado Postal J-48, Puebla 72570, Mexico}

\author{Kevin Peralta-Martinez}
\address{Instituto de F\'isica, Benem\'erita Universidad Aut\'onoma de Puebla, 
Apartado Postal J-48, Puebla 72570, Mexico}

\author{Jos\'e M. Sigarreta}
\address{Facultad de Matem\'aticas, Universidad Aut\'onoma de Guerrero, 
Carlos E. Adame No.54 Col. Garita, Acalpulco Gro. 39650, Mexico}

\author{Edson D. Leonel}
\address{Universidade Estadual Paulista (UNESP) - Departamento de F\'isica,
Av. 24A, 1515 -- Bela Vista -- CEP: 13506-900 -- Rio Claro -- SP -- Brazil}

\begin{abstract}
In this work we characterize the escape of orbits from the phase space of the Riemann-Liouville (RL) 
fractional standard map (fSM). The RL-fSM, given in action-angle variables, is derived from the 
equation of motion of the kicked rotor when the second order derivative is substituted by a RL 
derivative of fractional order $\alpha$.
Thus, the RL-fSM is parameterized by $K$ and $\alpha\in(1,2]$ which control the strength of nonlinearity 
and the fractional order of the  RL derivative, respectively. 
Indeed, for $\alpha=2$ and given initial conditions, the RL-fSM reproduces Chirikov's standard map.
By computing the survival probability $P_{\text{S}}(n)$ and the frequency of escape $P_{\text{E}}(n)$, 
for a hole of hight $h$ placed in the action axis, we observe two scenarios:
When the phase space is ergodic, both scattering functions are scale invariant with the typical escape 
time $n_{\text{typ}}=\exp\langle \ln n \rangle \propto (h/K)^2$.
In contrast, when the phase space is not ergodic, the scattering functions show a clear non-universal
and parameter-dependent behavior.
\end{abstract}

\end{frontmatter}



\section{Introduction}

When discussing about the generic transition to chaos, in the context of the Kolmogorov--Arnold--Moser 
(KAM) theorem, probably one of the most popular example models is the kicked rotor (KR); see e.g.~\cite{O08}.
The KR, which represents a free rotating stick in an inhomogeneous field that is periodically switched on 
in instantaneous pulses, is described by the second order differential equation
\begin{equation}
    \ddot{x} + K \sin(x) \sum_{n = 0}^\infty \delta\left(\frac{t}{T} - n \right) = 0 .
    \label{KR}
\end{equation}
Here, $x\in [0,2\pi]$ is the angular position of the stick, $K$ is the kicking strength, $T$ is the kicking 
period (that we set to one from now on), and $\delta$ is Dirac's delta function.
A very useful approach to the dynamics of the KR is by studying its stroboscopic projection, which is
well known as Chirikov's standard map (CSM)~\cite{C69}:
\begin{equation}
\begin{array}{ll}
p_{n+1} = p_n - K\sin(x_n) , \\
x_{n+1} = x_n + p_{n+1} , \qquad \mbox{mod}~(2\pi) ,\\
\end{array}
\label{CSM}
\end{equation}
where $p$ corresponds to the angular momentum of the KR's stick.
Indeed, CSM is known to represent the local dynamics of several Hamiltonian systems and is by itself 
a paradigm model of the KAM scenario.

In order to account for dynamical features not present in KAM's scenario, modified versions of CSM have 
been introduced. Among them we can highlight the dissipative version of CSM (also known as Zaslavsky 
map)~\cite{Z78} and the discontinuous version of CSM~\cite{B98}. 
Moreover, by substituting the second-order derivative in the equation of the KR by the Riemann-Liouville 
(RL) derivative $_0D_t^{\alpha}$, the RL fractional KR is obtained~\cite{TZ08,ET09}:
\begin{equation}
_0D_t^\alpha x + K \sin(x) \sum_{n = 0}^\infty \delta\left(t - n \right) = 0 , \quad 1< \alpha \leq 2 .
\label{fKR}
\end{equation}
Above
\begin{align}
&_0D_t^\alpha x(t) =  D_t^m {_0 I}_t^{m-\alpha}x(t) \nonumber \\ 
& = \frac{1}{\Gamma(m-\alpha)}\frac{d^{m}}{dt^{m}}\int_{0}^{t}\frac{x^{\tau}d\tau}{(t-\tau)^{\alpha-m+1}}, \quad m-1<\alpha\leq m, \nonumber
\end{align}
with $D_t^m = d^m/dt^m$ and $_0I_t^m f(t)$ is a fractional integral given by
\begin{equation*}
_0I_t^m f(t) = \frac{1}{\Gamma(m)}\int_0^t (t-\tau)^{\alpha-1}f(\tau)d\tau .
\end{equation*}
Notice that now $p(t) \equiv {_0D}_t^{\alpha-1}x(t)$.

Correspondingly, the stroboscopic projection of the RL fractional KR is known as the 
RL fractional standard map (RL-fSM) which reads as~\cite{ET09}
\begin{equation}
\begin{array}{ll}
p_{n+1} = p_n - K\sin(x_n) , \\
x_{n+1} = \displaystyle{ \frac{1}{\Gamma(\alpha)}\sum_{i=0}^n p_{i+1} V_\alpha(n-i+1) }  \\
	\qquad \quad \displaystyle{+ \frac{b}{\Gamma(\alpha-1)}(n+1)^{\alpha-2}, \quad \mbox{mod}~(2\pi). }
\end{array}
\label{RLfCSM}
\end{equation}
Here, $\Gamma$ is the Gamma function and 
\begin{equation*}
	V_\alpha(m) = m^{\alpha-1}-(m-1)^{\alpha-1}.
\end{equation*}
Note that the sum in the equation for the position in map (\ref{RLfCSM}) makes the RL-fSM to have memory, 
meaning that the future $(n+1)$--state depends on the entire orbit and not on the present $n$--state only.
The property of memory is known to be present in maps derived from fractional differential 
equations~\cite{T11b}, such as the RL-fSM of Eq.~(\ref{RLfCSM}), but also in maps derived from fractional 
integral equations~\cite{T21b} and in maps derived from fractional integro-differential equations~\cite{T21d}.

\begin{figure*}[ht]
\centering
\includegraphics[width=0.95\columnwidth]{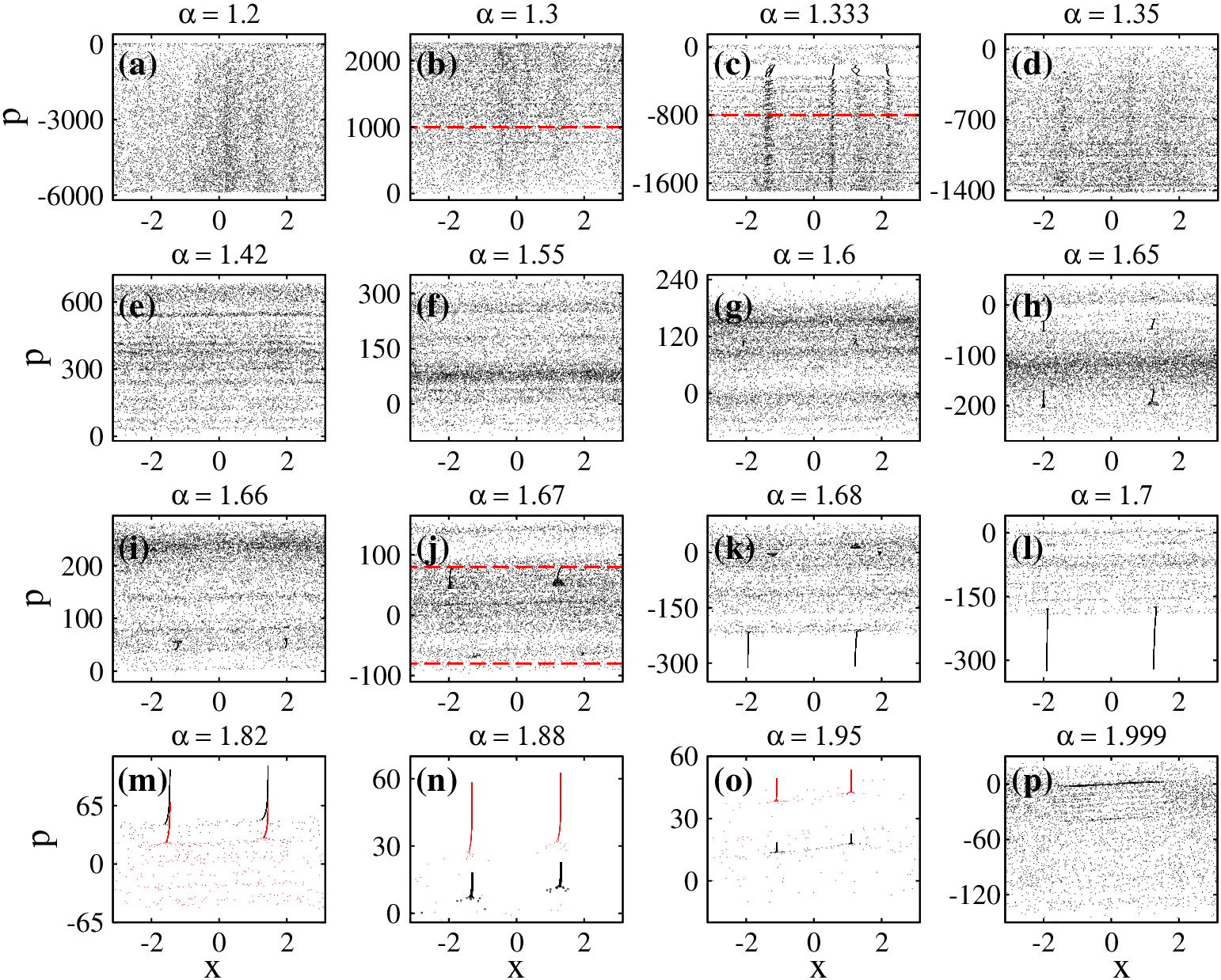}
\caption{Poincar\'e surfaces of section for the Riemann-Liouville fractional standard map of 
Eq.~(\ref{RLfSM}) with $K = 4.6$ and several values of $\alpha$. A single initial condition with 
$x_0=0$ and a random $p_0\in (0,2\pi/100)$ was iterated $10^4$ times (black dots). 
Red dots in panels (m,n,o) represent the orbit of a second initial condition.
Holes in the action axis in panels (b,c,j) are located at $p=\pm h$ (red dashed lines) with 
$h = 1000$, 800 and 80, respectively.}
\label{Fig01}
\end{figure*}

It is important to add that for $b=0$, in the case $\alpha = 2$, the RL-fSM reproduces CSM. 
Therefore in this work we consider the RL-fSM in the form
\begin{equation}
\begin{array}{ll}
p_{n+1} = p_n - K\sin(x_n) , \\
x_{n+1} = \displaystyle{ \frac{1}{\Gamma(\alpha)}\sum_{i = 0}^{n} p_{i+1} V_\alpha(n-i+1)}, \quad \mbox{mod}~(2\pi),
\end{array}
\label{RLfSM}
\end{equation}
where $1<\alpha \leq 2$ is assumed. The RL-fSM has no periodicity in $p$ and cannot be considered 
on a torus like CSM.

In contrast with CSM, depending on the strength of nonlinearity $K$ and the fractional order of the RL 
derivative $\alpha$, the RL-fSM generates attractors (fixed points, asymptotically stable periodic trajectories, 
slow converging and slow diverging trajectories, ballistic trajectories, and fractal-like structures) and/or chaotic 
trajectories~\cite{ET09,ET13,E19}.
Moreover, trajectories may intersect and attractors may overlap~\cite{E11}.
As an example, in Fig.~\ref{Fig01} we present Poincar\'e surfaces of section for the RL-fSM with $K = 4.6$ and 
several values of $\alpha$. 
In this figure we can observe the convergence to asymptotically stable periodic trajectories (see the period-two 
periodic orbits in panels (k) to (o)), the existence of unstable periodic trajectories (see the period-four periodic 
orbit in panel (c)), as well as quite uniform chaotic trajectories (see panels (a) and (d)).
Similar Poincar\'e surfaces of section are observed for other values of $K$.
Thus, we can safely state that the RL-fSM is a richer dynamical system as compared to CSM.

Since its investigation by Edelman and Tarasov~\cite{ET09}, several studies on the RL-fSM have been reported
and other fractional versions of the standard map have also been introduced, see 
e.~g.~\cite{T11b,E11,E13,TE10,T11,T13,E14,T21c,EH22}.
Among those studies we highlight that:
Already in Refs.~\cite{TE10,T11} the dissipative fSM (which is the fractional version of
Zaslavsky map~\cite{Z78}) was introduced;
the Caputo fSM (i.e.~the fSM derived from the KR with a Caputo fractional derivative) was introduced and 
contrasted with the RL-fSM in Ref.~\cite{E11};
the $\alpha$-family of the fSM, where the order of the fractional derivative is not restricted to $1<\alpha \leq 2$,
was defined in Refs.~\cite{E13,E14,EH22}.

Coming back to the RL-fSM, even though many of its properties have been already studied, as far as we know, 
its scattering properties have not been explored yet. Thus, in this work we undertake this task and characterize 
the escape of orbits from the phase space of the RL-fSM.
Specifically, in this work we compute the survival probability $P_{\text{S}}(n)$ and the frequency of 
escape $P_{\text{E}}(n)$ for a hole of hight $h$ placed in the action axis. Then, we analyze both scattering 
quantities as a function of $K$ and $\alpha$, the parameters of the RL-fSM.

\section{Scattering setup and scattering measures}

Scattering experiments are used to probe the properties of target systems by measuring
transport or scattering quantities~\cite{APT13}.
In classical scattering we can cite two main scattering setups: 
in one setup particles are measured after they are scattered by a target system while in another 
setup the target system is characterized by means of particle leaking.
Here we use the second setup where the leak can be a
physical hole (such as an opening in the boundary of a billiard table)
or a subset of the phase space (i.e.,~a threshold in one or more variables).
In either case, an orbit with initial conditions inside the dynamical system that
reaches the hole is considered to escape from the system, see e.g.~\cite{ODCL13,MMLL15,OPML21}. 
In studies of classical leaking, the survival probability $P_{S}(n)$ and the frequency of escape $P_{E}(n)$
are widely used quantities; hence, in this work we compute both.

We open the RL-fSM by placing a hole on a subset of the phase space of constant action. 
Furthermore, we set the hole as two horizontal lines at $p=\pm h$ in the Poincar\'e surfaces of section; 
this is due to the symmetry of the phase space around $p=0$, i.e.~$\langle p_n \rangle = 0$. In Figs.~\ref{Fig01}(b), 
\ref{Fig01}(c) and~\ref{Fig01}(j) we show examples of holes with $h = 1000$, 800 and 80, respectively (see red 
dashed lines). Thus, controlling the nonlinearity ($K$) of map (\ref{RLfSM}), the fractional order of the RL 
derivative ($\alpha$), and the openness of the scattering setup ($h$), we analyze the escape of trajectories from 
the  RL-fSM.

For each combination of $(K,h,\alpha)$, we consider an ensemble of $10^5$ orbits
having as initial conditions $x_0=0$ and random $p_0$ uniformly distributed in the interval $(0,2\pi/100)$.
Trajectories with each initial condition evolve in time according to map~(\ref{RLfSM}).
Once $|p_n|>h$, we conclude that the orbit has escaped from the RL-fSM and choose a new initial condition.
We count the number of orbits $N_{\text{S}}(n)$, that at time $n$, have not escaped yet
from the RL-fSM, then we compute $P_{\text{S}}(n)$ as $P_{\text{S}}(n) = N_{\text{S}}(n)/M$.
While keeping track of the total number of orbits that have escaped up to time $n$, we simultaneously 
construct a histogram for the frequency of escape $P_{\text{E}}(n)$.

\section{Survival probability and frequency of escape}

Since the RL-fSM displays a rich and diverse dynamics we characterize the orbits leaking 
from its phase space first when the phase space is ergodic and later when it is not.
Also, without loss of generality, in the following numerical calculations we consider two relatively 
large values of $K$ (6.908745 and 7.5) which allowed us to get good statistics in reasonable
computer times. We note that some aspects of the dynamics of 
the RL-fSM with $K=6.908745$ have been explored in~\cite{ET09}, so we decided to use this
value of $K$ in our study.
Moreover, we verified that our conclusions do not depend on these values
of $K$.

\subsection{Ergodic phase space}

\begin{figure}
\centering
\includegraphics[width=0.7\columnwidth]{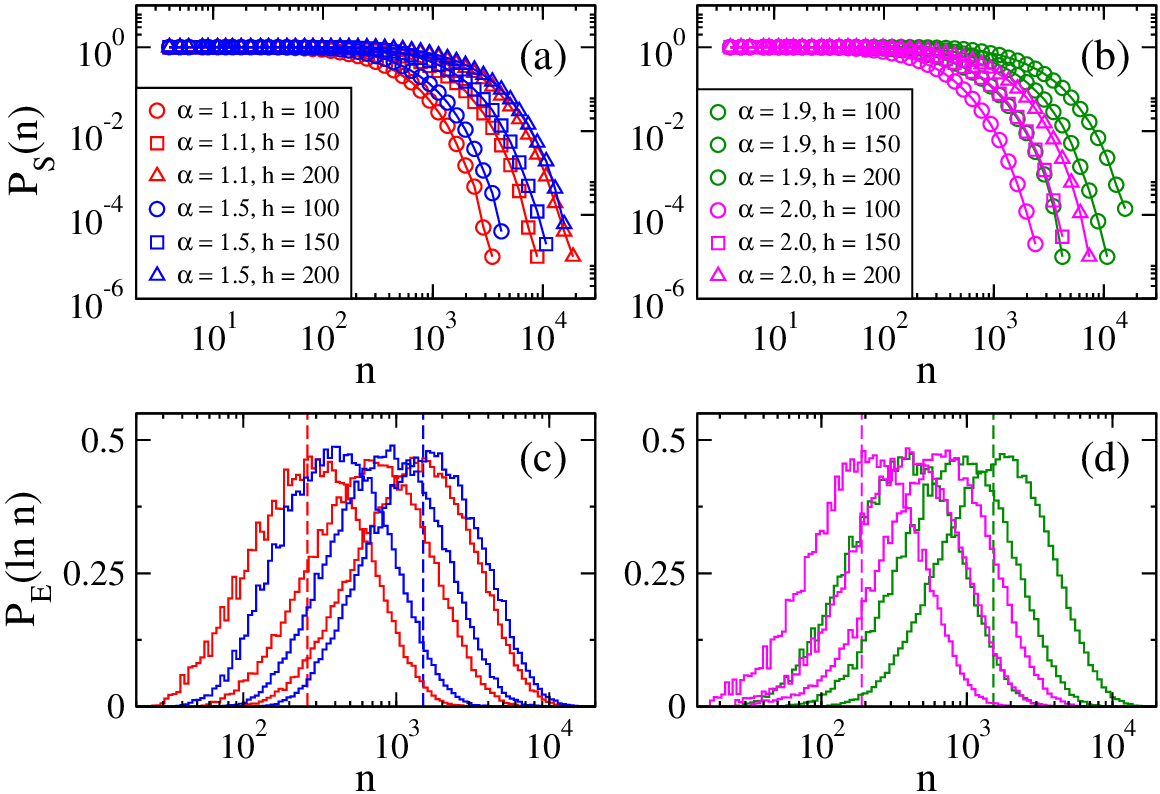}
\caption{(a,b) Survival probability $P_{\text{S}}(n)$ as a function of $n$ (symbols) for the RL-fSM with 
$K = 6.908745$ and some combinations of $\alpha$ and $h$
($\alpha=1.1$ in red, $\alpha=1.5$ in blue, $\alpha=1.9$ in green, and $\alpha=2$ in magenta; $h$ grows 
from left to right, as labeled in the figure). 
Full lines are fittings of the data with Eq.~(\ref{PS}).
Curves were computed from $M = 10^5$ orbits.
(c,d) Histograms for the frequency of escape $P_{\text{E}}(\ln n)$. Same color code as in (a,b). 
The typical escape times $n_{\text{typ}}=\exp\langle \ln n \rangle$ are indicated by the vertical dashed lines (they are shown
for four histograms only to avoid figure saturation).
Histograms were computed from $M = 10^5$ orbits.}
\label{Fig02}
\end{figure}

In Figs.~\ref{Fig02}(a,b) we present the survival probability $P_{\text{S}}(n)$ as a function of $n$ for the 
RL-fSM with $K = 6.908745$ and several combinations of $\alpha$ and $h$. 
The values of $\alpha$ we choose in Figs.~\ref{Fig02}(a,b) produce an ergodic phase space, 
see Fig.~\ref{FigA01}; or at least we did not observe the formation of structures in phase space for the
values of $h$ we consider. Note that we are considering here the value of $\alpha=2$, see the magenta 
curves in Fig.~\ref{Fig02}(b), which corresponds to CSM whose scattering properties have already
been studied in~\cite{MMLL15,OPML21}.
From Figs.~\ref{Fig02}(a,b) we note that $P_{\text{S}}(n)$ shows an almost perfect exponential 
decay of the form
\begin{equation}
P_{\text{S}}(n) = \exp\left(-\frac{n}{\mu}\right) ,
\label{PS}
\end{equation}
which is typical of strongly chaotic systems~\cite{APT13,SODB93} (see also~\cite{MMLL15,OPML21}).
Indeed, the full lines in Figs.~\ref{Fig02}(a,b) are fittings of Eq.~(\ref{PS}) to the data (symbols), were
$\mu$ has been used as a fitting parameter.

For the frequency of escape $P_{\text{E}}(n)$, we found that it grows with $n$, reaches a 
maximum value, and then decreases to zero for exponentially large iteration times. Thus, in order to clearly 
observe the complete panorama, we compute $P_{\text{E}}(\ln n)$ instead (see also~\cite{MMLL15,OPML21}).
 Therefore, in Figs.~\ref{Fig02}(c,d) we present 
$P_{\text{E}}(\ln n)$ for for the RL-fSM with $K = 6.908745$ and the same combinations of $\alpha$ 
and $h$ used in panels (a,b) for $P_{\text{S}}(n)$.

\begin{figure}
\centering
\includegraphics[width=0.7\columnwidth]{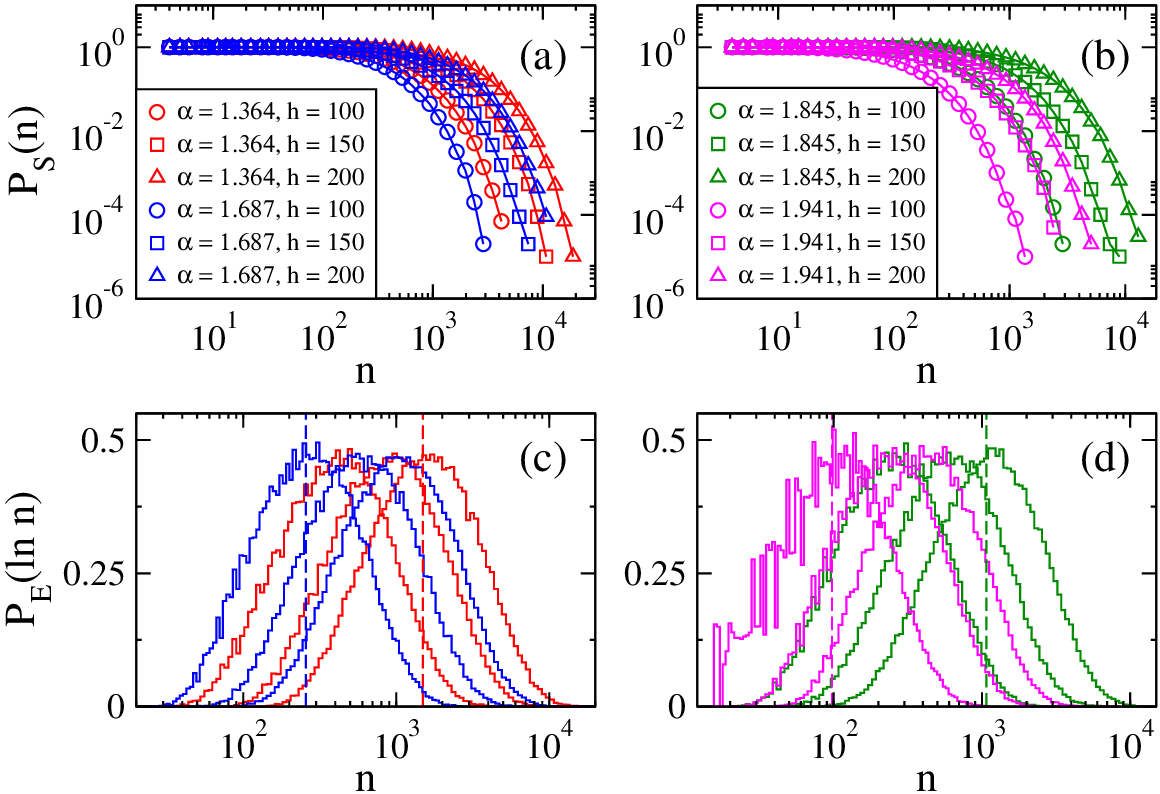}
\caption{(a,b) Survival probability $P_{\text{S}}(n)$ as a function of $n$ (symbols) for the RL-fSM with 
$K = 7.5$ and some combinations of $\alpha$ and $h$
($\alpha=1.364$ in red, $\alpha=1.687$ in blue, $\alpha=1.845$ in green, and $\alpha=1.941$ in magenta; $h$ grows 
from left to right, as labeled in the figure). 
Full lines are fittings of the data with Eq.~(\ref{PS}).
Curves were computed from $M=10^5$ orbits.
(c,d) Histograms for the frequency of escape $P_{\text{E}}(\ln n)$. Same color code as in (a,b). 
The typical escape times $n_{\text{typ}}=\exp\langle \ln n \rangle$ are indicated by the vertical dashed lines  (they are shown
for four histograms only to avoid figure saturation).
Histograms were computed from $M=10^{5}$ orbits.}
\label{Fig03}
\end{figure}

To stress the fact that the panorama shown in Fig.~\ref{Fig02} is generic for the RL-fSM when
the parameters $K$ and $\alpha$ produce an ergodic phase space, in Fig.~\ref{Fig03} we show
$P_{\text{S}}(n)$ and $P_{\text{E}}(\ln n)$ now for $K=7.5$ and several different combinations of 
$\alpha$ and $h$.
In Fig.~\ref{FigA02} we present the Poincar\'e surfaces of section corresponding to the chosen 
values of $\alpha$.

\begin{figure}
\centering
\includegraphics[width=0.7\columnwidth]{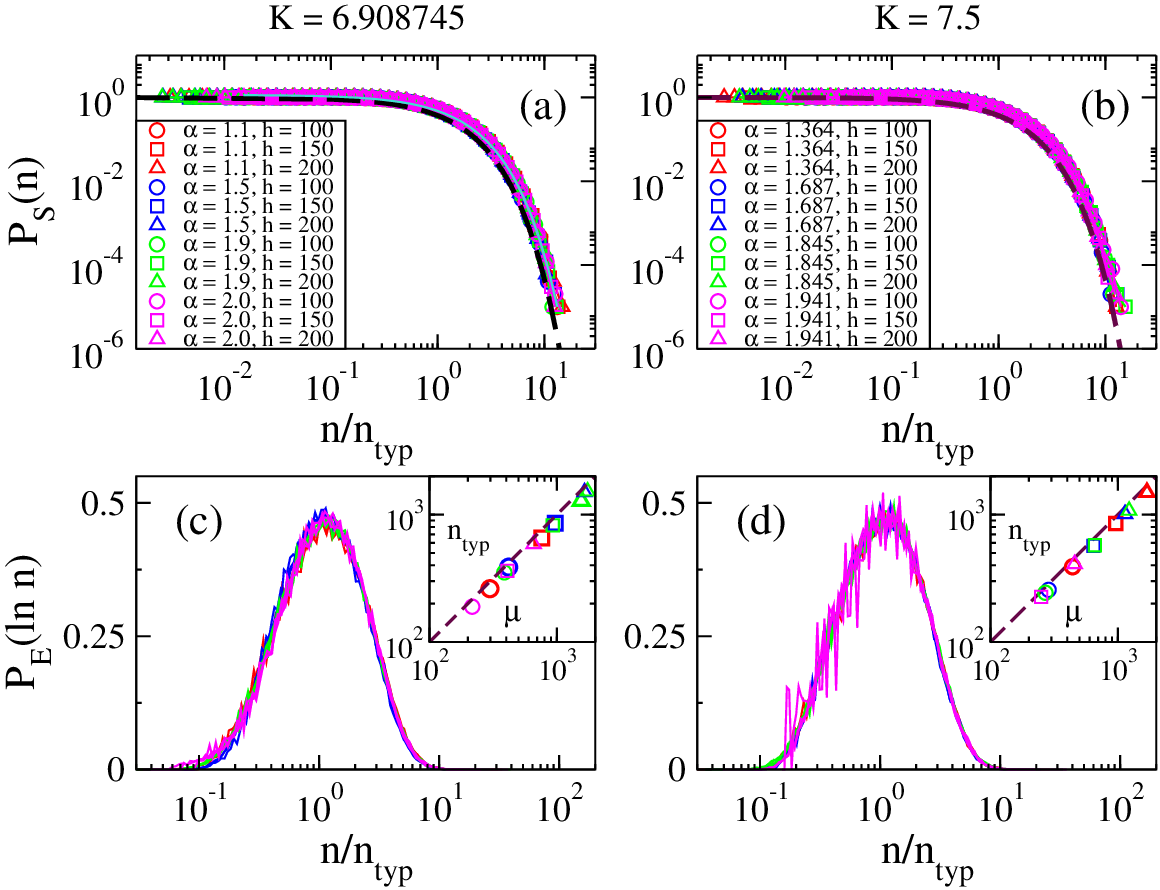}
\caption{Survival probability $P_{\text{S}}(n)$ as a function of $n/n_{\text{typ}}$ 
for the RL-fSM with (a) $K = 6.908745$ and (b) $K=7.5$; same curves of (a) Figs.~\ref{Fig02}(a,b) and (b) 
Figs.~\ref{Fig03}(a,b). Dashed black lines in (a,b) correspond to Eq.~(\ref{PSntyp}).
Insets in (c,d) show $n_{\text{typ}}$ vs.~$\mu$. 
The values of $\mu$ were extracted by fitting Eq.~(\ref{PS}) to the $P_{\text{S}}(n)$ curves of (a) 
Figs.~\ref{Fig02}(a,b) and (b) Figs.~\ref{Fig03}(a,b).
The relation $n_{\text{typ}}=\mu$ (dashed lines) is shown as a reference.
Histograms for the frequency of escape $P_{\text{E}}(\ln n)$ as a function of 
$n/n_{\text{typ}}$ for the RL-fSM with (c) 
$K = 6.908745$ and (d) $K=7.5$; same curves of (c) Fig.~\ref{Fig02}(c,d) and (d) Fig.~\ref{Fig03}(c,d).}
\label{Fig04}
\end{figure}

In~\cite{OPML21}, it was shown that the {\it typical iteration time}, 
\begin{equation}
n_{\text{typ}} = \exp{\left\langle \ln n \right\rangle} ,
\label{ntypical}
\end{equation}
characterizes well the maximum of $P_{\text{E}}(\ln n)$ of strongly chaotic systems. 
This fact is also valid for the RL-fSM with an ergodic phase space, as can be seen in 
Figs.~\ref{Fig02}(c,d) and Figs.~\ref{Fig03}(c,d) where 
$n_{\text{typ}}$ is indicated with vertical dashed lines for selected histograms. 
Moreover, the relation between $P_{\text{E}}(n)$ and $P_{\text{S}}(n)$,
\begin{equation}
P_{\text{E}}(n) = - \frac{d P_{\text{S}}(n)}{dn} ,
\label{PEPS}
\end{equation}
let us state that $\mu = n_{\text{typ}}$ and, accordingly, allows us to write
\begin{equation}
P_{\text{S}}(n) \approx \exp\left(-\frac{n}{n_{\text{typ}}}\right) .
\label{PSntyp}
\end{equation}
Note that in Eq.~(\ref{PSntyp}) we are writing approximately equal instead of
equal. This is because we numerically found that $\mu \approx n_{\text{typ}}$, as can be
seen in the insets of Figs.~\ref{Fig04}(a,b) where we plot $n_{\text{typ}}$ vs.~$\mu$; here
$\mu$ is extracted from the fittings of Eq.~(\ref{PS}) to the $P_{\text{S}}(n)$ curves of
Figs.~\ref{Fig02}(a,b) and Figs.~\ref{Fig03}(a,b).

\begin{figure}
\centering
\includegraphics[width=\columnwidth]{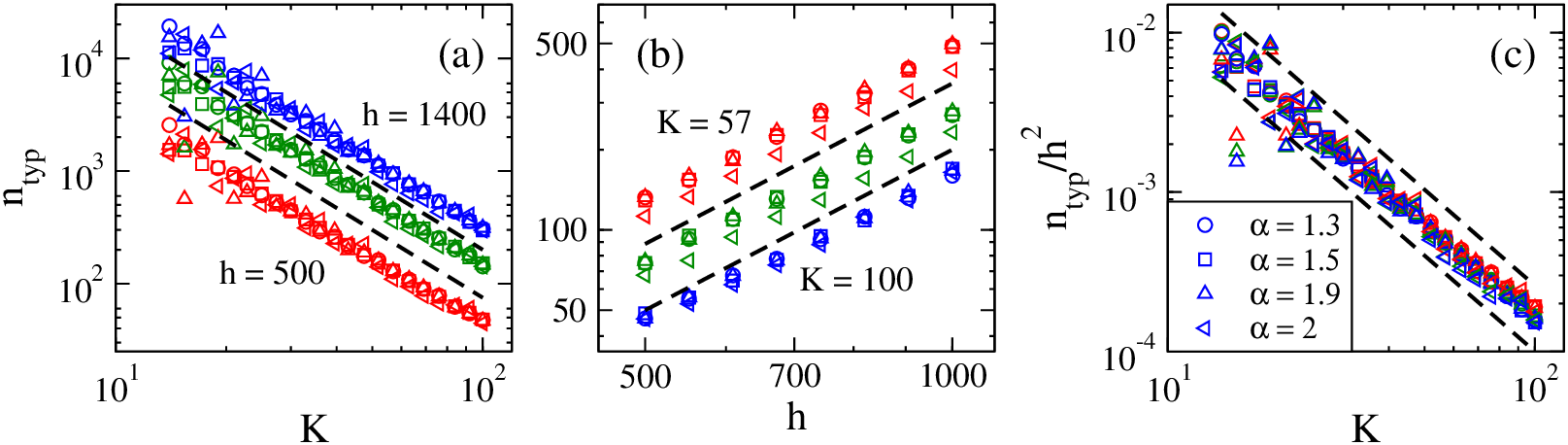}
\caption{(a) Typical iteration time $n_{\text{typ}}$ as a function of $K$ for the RL-fSM. 
Here $h=500$ (red symbols), $h=950$ (green symbols), and $h=1400$ (blue symbols);
different values of $\alpha$ are shown: $\alpha=1.3$ ($\circ$), $\alpha=1.5$ ($\square$), 
$\alpha=1.9$ ($\triangle$), and $\alpha=2$ ($\triangleleft$).
$n_{\text{typ}}\propto K^{-2}$ (dashed lines) is shown as a reference.
(b) $n_{\text{typ}}$ as a function of $h$. 
$K=57$ (red symbols), $K=76$ (green symbols), and $K=100$ (blue symbols) for the values 
of $\alpha$ reported in (a). 
$n_{\text{typ}}\propto  h^2$ (dashed lines) is shown as a reference.
(c) Typical escape time $n_{\text{typ}}$, normalized to $h^2$, as a function of $K$.  
Same data as in panel (a). 
$n_{\text{typ}}\propto K^{-2}$ (dashed lines) is shown as a reference.}
\label{Fig05}
\end{figure}

Expression~(\ref{PSntyp}) is of paramount importance 
because it reveals that $P_{\text{S}}(n)$ and accordingly $P_{\text{E}}(\ln n)$ depend
on the ratio $n/n_{\text{typ}}$ only; meaning that the typical iteration time (see Eq.~(\ref{ntypical}))
is the scaling parameter of both quantities.
In practical terms, this means that when plotting $P_{\text{S}}(n)$ vs.~$n/n_{\text{typ}}$ and $P_{\text{E}}(\ln n)$ 
vs.~$n/n_{\text{typ}}$, all curves will fall on top of universal curves
independently of the parameter combination $(K,\alpha,h)$.
Indeed, we verify this last statement in Figs.~\ref{Fig04}(a,b) and Figs.~\ref{Fig04}(c,d) where we
present $P_{\text{S}}(n)$ and $P_{\text{E}}(\ln n)$ as a function of $n/n_{\text{typ}}$, respectively, for 
the RL-fSM and several parameter combinations (in fact we are using the same curves reported in
Figs.~\ref{Fig02} and~\ref{Fig03}). As a reference, we are also including 
a plot of Eq.~(\ref{PSntyp}) in Figs.~\ref{Fig04}(a,b) as black dashed lines to confirm that it describes relatively
well the corresponding numerical data.

Finally, given the relevance of the typical iteration time for the scattering quantities we study here, it is 
useful to look for the dependence of $n_{\text{typ}}$ on the system parameters $(K,\alpha,h)$.
Then, in Fig.~\ref{Fig05}(a), we plot $n_{\text{typ}}$ as a function of $K$ for several combinations
of $\alpha$ and $h$, and in Fig.~\ref{Fig05}(b), we report $n_{\text{typ}}$ as a function of $h$ for 
several combinations of $\alpha$ and $K$. 
Since from these figures we observe that $n_{\text{typ}}$ depends 
on both $K$ and $h$ as power-laws, while there is not an evident dependence on $\alpha$, we propose 
the following scaling hypotheses for $n_{\text{typ}}$:
\begin{equation}
n_{\text{typ}} \propto K^{\gamma_K}h^{\gamma_h} .
\label{ntypscaling}
\end{equation}
where $\gamma_K$ and $\gamma_h$ are scaling exponents.
By performing fittings of the data of Figs.~\ref{Fig05}(a,b) with Eq.~(\ref{ntypscaling}) we obtained
$\gamma_K \approx -2$ and $\gamma_h \approx 2$, see the dashed lines in Figs.~\ref{Fig05}(a,b). 
Indeed, by plotting now $n_{\text{typ}}/h^2$ vs.~$K$, see Fig.~\ref{Fig05}(c), we better observe
that $n_{\text{typ}} \propto K^{-2}$ and confirm the independence of $n_{\text{typ}}$ on $\alpha$.
It is important to mention that 
(i) the values of $\gamma_K$ and $\gamma_h$ we found here for the
RL-fSM were also reported for the discontinuous standard map in the quasilinear diffusion 
regime~\cite{OPML21} and
(ii) the behavior $n_{\text{typ}}\propto K^{-2}$ is clearly observed for relatively large values of $K$ 
only, i.e.~$K\stackrel{>}{\sim} 30$; see Figs.~\ref{Fig05}(a,c).

\subsection{Non-ergodic phase space}

Once we verify that the escape of orbits from the phase space of the RL-fSM, when characterized 
by an ergodic phase space, is similar to that reported for strongly chaotic systems, we compute
the survival probability and the histograms for the frequency of escape when the phase
space is non-ergodic. We again consider the RL-fSM with $K = 6.908745$ and $K=7.5$, as in the
previous subsection, but now we choose values of $\alpha$ which produce non-ergodic Poincar\'e 
surfaces of section, see Fig.~\ref{FigA03}.

\begin{figure}
\centering
\includegraphics[width=0.7\columnwidth]{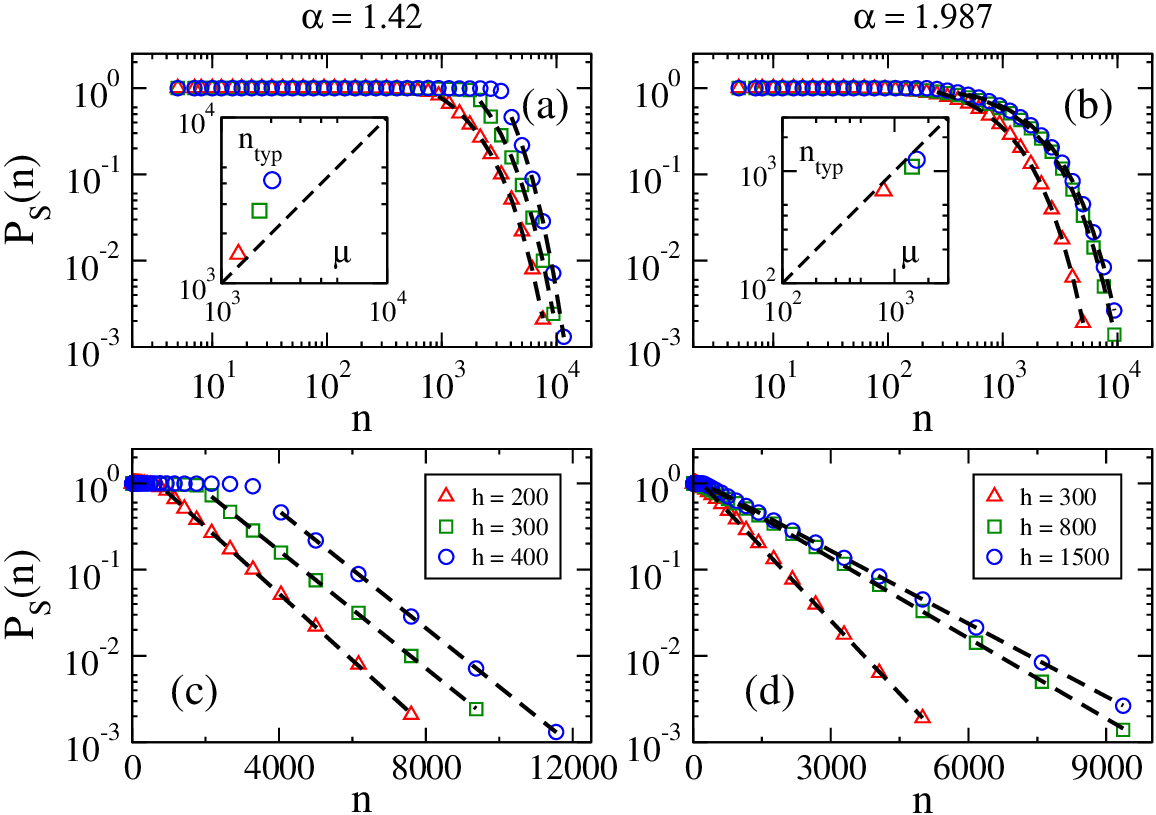} 
\caption{
Survival probability $P_{\text{S}}(n)$ as a function of $n$ (symbols) for the RL-fSM with 
$K = 6.908745$ and some combinations of $\alpha$ and $h$: 
(a,c) $\alpha=1.42$ and (b,d) $\alpha=1.987$; 
$h$ grows from left to right, as labeled in the figure. 
Upper and lower panels contain the same curves, however the horizontal axes
in the lower panels are in linear scale to better appreciate the exponential decay of $P_{\text{S}}(n)$.
Dashed black lines are fittings of the data with Eq.~(\ref{PS}).
The $R^2-1$ values of the fittings are: 
(a,c) $1.2\times 10^{-4}$ ($h=200$), $1.4\times 10^{-4}$ ($h=300$), $4.4\times 10^{-6}$ ($h=400$), and 
(b,d) $1.9\times 10^{-5}$ ($h=300$), $1.5\times 10^{-4}$ ($h=800$), $5.1\times 10^{-6}$ ($h=1500$).
Curves were computed from $M = 10^5$ orbits.
Insets in (a,b) show $n_{\text{typ}}$ vs.~$\mu$. 
The values of $\mu$ were extracted by fitting Eq. (6) to the $P_{\text{S}}(n)$ curves in 
the main panels.
The relation $n_{\text{typ}}=\mu$ (dashed lines) is shown as a reference.
}
\label{Fig06a}  
\end{figure}
\begin{figure}[h!]
\centering
\includegraphics[width=0.7\columnwidth]{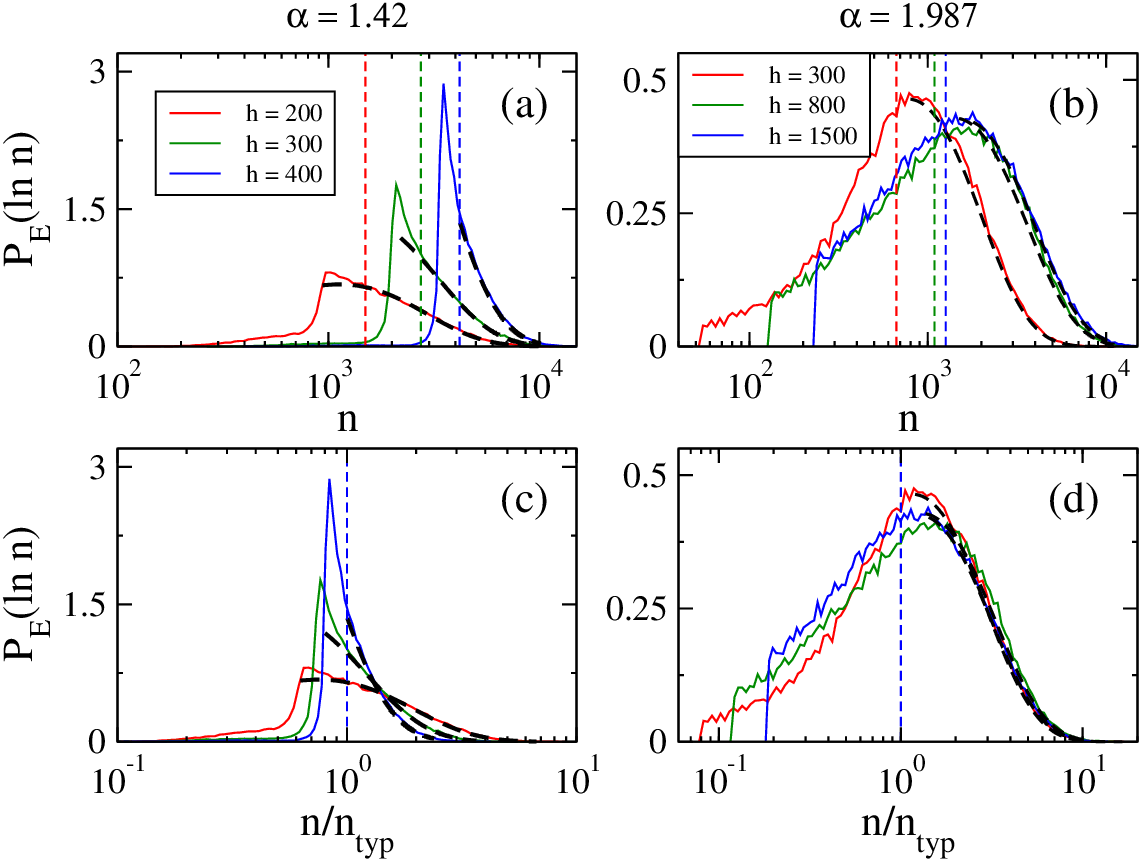} 
\caption{
(a,b) Histograms for the frequency of escape $P_{\text{E}}(\ln n)$. Same color code as in Fig.~\ref{Fig06a}. 
The typical escape times $n_{\text{typ}}=\exp\langle \ln n \rangle$ are indicated by the vertical dashed lines.
Histograms were computed from $M = 10^5$ orbits. Black dashed lines are Eq.~(\ref{PEmu}).
(c,d) $P_{\text{E}}(\ln n)$ as a function of $n/n_{\text{typ}}$; same histograms as in panels (a,b), respectively.
(a,c) $\alpha=1.42$ and (b,d) $\alpha=1.987$.
}
\label{Fig06b}  
\end{figure}
\begin{figure}
\centering
\includegraphics[width=0.7\columnwidth]{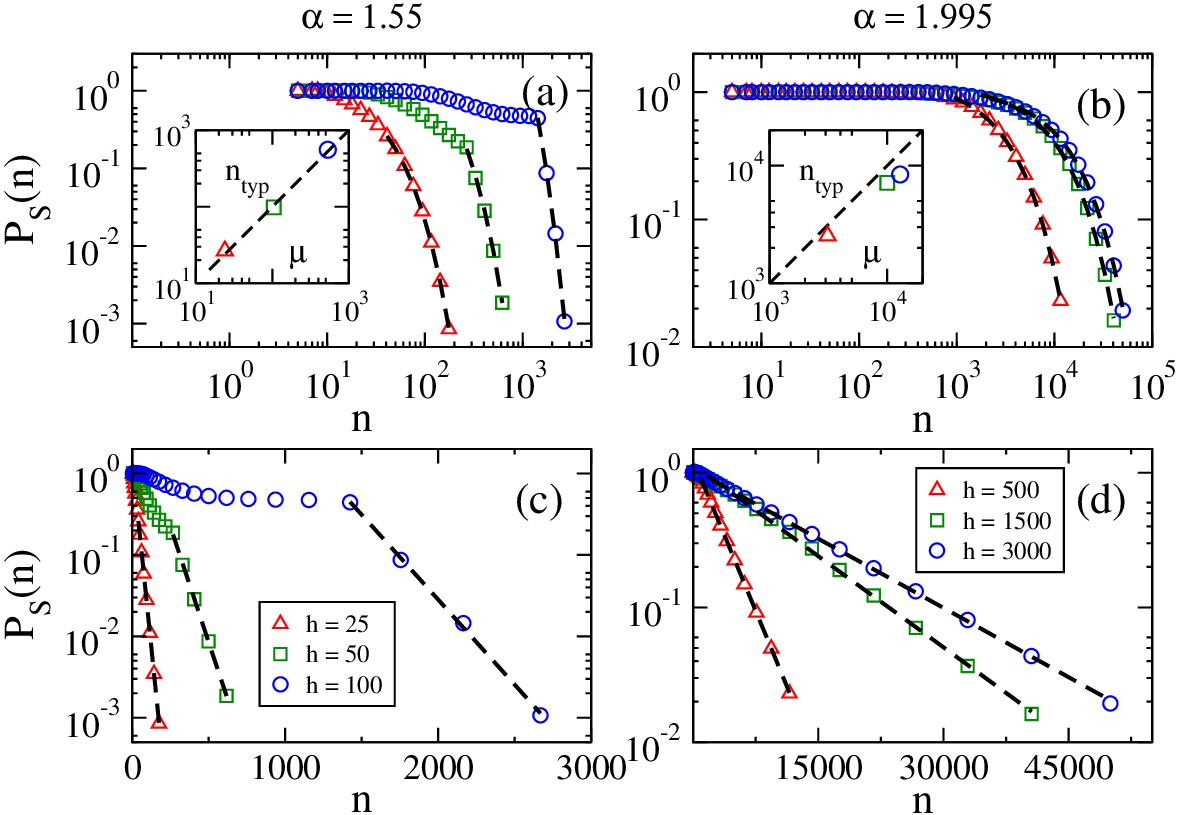}
\caption{
Survival probability $P_{\text{S}}(n)$ as a function of $n$ (symbols) for the RL-fSM with 
$K = 7.5$ and some combinations of $\alpha$ and $h$: 
(a,c) $\alpha=1.55$ and (b,d) $\alpha=1.995$; $h$ grows from left to right, as labeled in the figure. 
Upper and lower panels contain the same curves, however the horizontal axes
in the lower panels are in linear scale to better appreciate the exponential decay of $P_{\text{S}}(n)$.
Dashed black lines are fittings of the data with Eq.~(\ref{PS}).
The $R^2-1$ values of the fittings are: 
(a,c) $2.1\times 10^{-5}$ ($h=25$), $4.2\times 10^{-4}$ ($h=50$), $1.1\times 10^{-3}$ ($h=100$), and 
(b,d) $1.6\times 10^{-5}$ ($h=500$), $6.5\times 10^{-5}$ ($h=1500$), $3.4\times 10^{-4}$ ($h=3000$).
Curves were computed from $M = 10^5$ orbits.
Insets in (a,b) show $n_{\text{typ}}$ vs.~$\mu$. 
The values of $\mu$ were extracted by fitting Eq. (6) to the $P_{\text{S}}(n)$ curves in 
the main panels.
The relation $n_{\text{typ}}=\mu$ (dashed lines) is shown as a reference.
}
\label{Fig07a}
\end{figure}
\begin{figure}[h!]
\centering
\includegraphics[width=0.7\columnwidth]{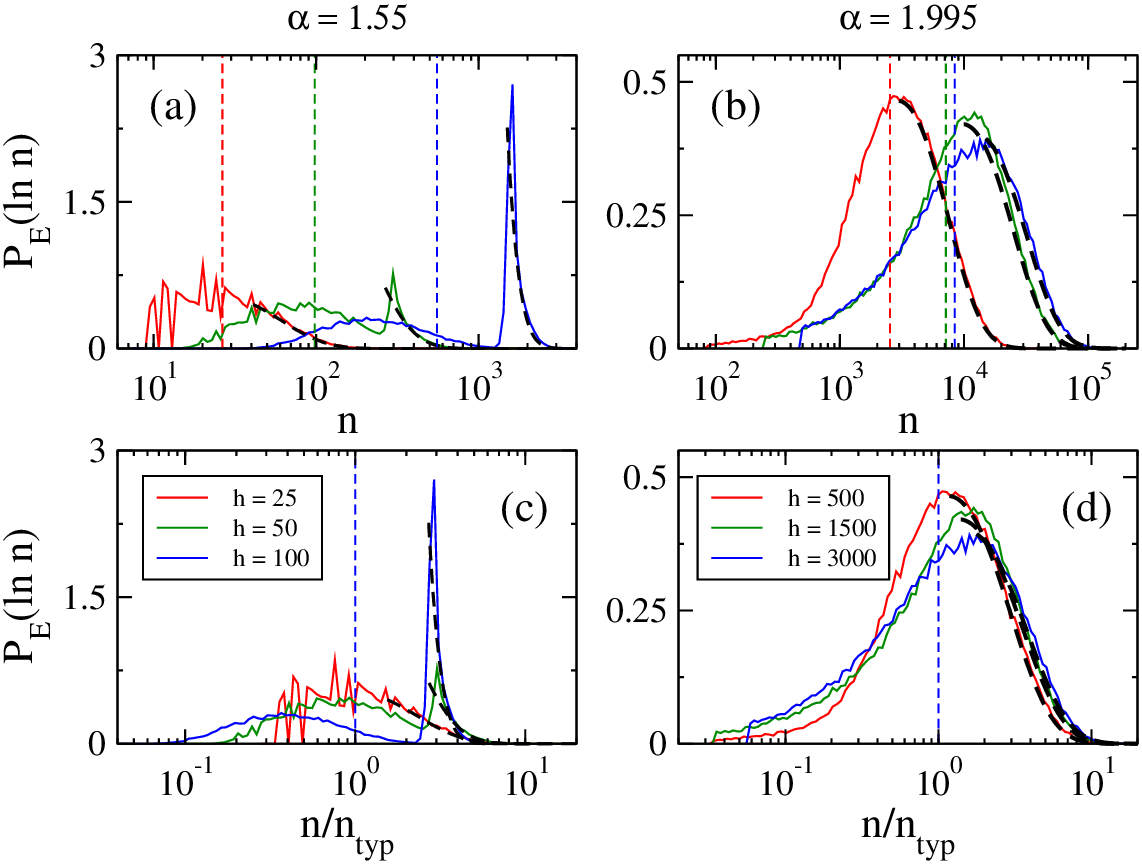}
\caption{
(a,b) Histograms for the frequency of escape $P_{\text{E}}(\ln n)$. Same color code as in Fig.~\ref{Fig07a}.  
The typical escape times $n_{\text{typ}}=\exp\langle \ln n \rangle$ are indicated by the vertical dashed lines.
Histograms were computed from $M = 10^5$ orbits.
Black dashed lines are Eq.~(\ref{PEmu}).
(c,d) $P_{\text{E}}(\ln n)$ as a function of $n/n_{\text{typ}}$; same histograms as in panels (a,b), respectively.
(a,c) $\alpha=1.55$ and (b,d) $\alpha=1.995$.
}
\label{Fig07b}
\end{figure}

Then, in Fig.~\ref{Fig06a}, we present $P_{\text{S}}(n)$ as a function of $n$ for the RL-fSM 
with $K = 6.908745$ and two values of $\alpha$ [(a,c) $\alpha=1.42$ and (b,d) $\alpha=1.987$] for some 
values of $h$. In Fig.~\ref{Fig06b}(a,b), we plot the corresponding $P_{\text{E}}(\ln n)$ histograms.

For increasing $h$ both functions $P_{\text{S}}(n)$ and $P_{\text{E}}(\ln n)$ are displaced to the 
right as in the ergodic phase space case, see Figs.~\ref{Fig02} and~\ref{Fig03}; this is expected 
since particles take longer to escape the higher the hole is. However, now $P_{\text{S}}(n)$ can 
not be described by a simple exponential decay, as in Eq.~(\ref{PS}).
We stress that not even the survival probability curves reported in Fig.~\ref{Fig06a}(b), which
look like simple exponentials, can be fitted by Eq.~(\ref{PS}) in the complete range.
This is not a big surprise since the exponential decay of $P_{\text{S}}(n)$ is only expected
for strongly chaotic systems~\cite{APT13,MMLL15,OPML21,SODB93}, which is not the case with 
the values of $\alpha$ we chose in Figs.~\ref{Fig06a} and~\ref{Fig06b} [see Fig.~\ref{FigA03}(a,b)]. 

In Figs.~\ref{Fig07a} and~\ref{Fig07b}, we present $P_{\text{S}}(n)$ and $P_{\text{E}}(\ln n)$,
respectively, for the RL-fSM 
now with $K = 7.5$ and (a,c) $\alpha=1.55$ and (b,d) $\alpha=1.995$. Here we also observe, 
as expected since the phase space is not ergodic [see Fig.~\ref{FigA03}(c,d)], that $P_{\text{S}}(n)$ 
and $P_{\text{E}}(\ln n)$ strongly depend on the map parameters.

Nevertheless we found that the decay of $P_{\text{S}}(n)$ in both cases, Figs.~\ref{Fig06a} 
and~\ref{Fig07a}, is indeed exponential for large times. See the black dashed lines in those
figures which are fittings of the $P_{\text{S}}(n)$ curves with Eq.~(\ref{PS}).
In the captions of Figs.~\ref{Fig06a} and~\ref{Fig07a}, we report the $R^2$ (coefficient 
of determination) values of the fittings which validate the exponential decay of $P_{\text{S}}(n)$; 
in fact, the values of $R^2$ are so close to one that we better report $R^2-1$.
Clearly, even when the tail of $P_{\text{S}}(n)$ can be fitted with Eq.~(\ref{PS}), the values of
$\mu$ extracted from the fittings can not always be identified with $n_{\text{typ}}$ as displayed in the
insets of Figs.~\ref{Fig06a}(a,b) and~\ref{Fig07a}(a,b).
Luckily, once we know that the tail of $P_{\text{S}}(n)$ in Figs.~\ref{Fig06a} 
and~\ref{Fig07a} is well described by Eq.~(\ref{PS}),
by the use of Eq.~(\ref{PEPS}) we can estimate the tail of $P_{\text{E}}(\ln n)$ as
\begin{equation}
P_{\text{E}}(\ln n) \sim \frac{n}{\mu} \exp\left(-\frac{n}{\mu}\right) .
\label{PEmu}
\end{equation}
Indeed, Eq.~(\ref{PEmu}) describes well the tail of $P_{\text{E}}(\ln n)$ for all the parameter
combinations reported in Figs.~\ref{Fig06b}(a,b) and~\ref{Fig07b}(a,b), see the dashed lines.

Finally, it is pertinent to mention that even when we were able to describe the tails of 
$P_{\text{S}}(n)$ and $P_{\text{E}}(\ln n)$ with Eqs.~(\ref{PS}) and~(\ref{PEmu}), respectively,
these scattering functions can not be scaled neither with $n_{\text{typ}}$ nor with $\mu$.
See for example the histograms of $P_{\text{E}}(\ln n)$ as a function of $n/n_{\text{typ}}$
in Figs.~\ref{Fig06b}(c,d) and~\ref{Fig07b}(c,d).

\section{Discussion and conclusions}

In this work, we characterized the leaking of orbits from the phase space of the Riemann-Liouville  
fractional standard map (RL-fSM). The RL-fSM is parameterized by $K$ and $\alpha\in(1,2]$ which 
control the strength of nonlinearity and the fractional order of the derivative of the corresponding
fractional kicked rotor. 
It is important to stress that, to the best of our knowledge, the scattering properties of maps with 
memory have not been explored before.

We computed the frequency of escape $P_{\text{E}}(n)$ and the survival probability $P_{\text{S}}(n)$, 
more specifically $P_{\text{S}}(\ln n)$, for a hole of hight $h$ placed in the action axis.
We explored two scenarios: one where the phase space of the RL-fSM is ergodic, see 
e.g.~Figs.~\ref{FigA01} and~\ref{FigA02}, and another where the phase space is non-ergodic, see 
e.g.~Fig.~\ref{FigA03}. 

When the phase space of the RL-fSM is ergodic we found that $P_{\text{E}}(n)$ and $P_{\text{S}}(\ln n)$
are both scale invariant with the typical escape time $n_{\text{typ}}=\exp\langle \ln n \rangle$, so they are 
well described by universal curves; see Fig.~\ref{Fig04}.
This is in agreement with previous studies on leaking form discontinuous maps and
dissipative maps~\cite{MMLL15,OPML21}.
Moreover, for strongly chaotic systems, it has been shown that the survival probability can be obtained
from the solution of the diffusion equation describing the transport of particles along the phase space 
as~\cite{ODCL13,MMLL15}
\begin{equation}
P_{\text{S}}(n) \approx \exp\left(-\frac{\pi^2D}{4h^2} n \right) ,
\label{PSD}
\end{equation}
where $D$ is the diffusion coefficient.
Therefore, by equating Eqs.~(\ref{PSntyp}) and~(\ref{PSD}) and with the help of Eq.~(\ref{ntypscaling}), 
we can infer the diffusion coefficient of the RL-fSM as
\begin{equation}
D \approx \frac{4}{\pi^2} \frac{h^2}{n_{\text{typ}}} \propto K^2 .
\label{D}
\end{equation}
It is relevant to highlight the independence of $D$ on $\alpha$.

When the phase space of the RL-fSM is not ergodic, even though we were able to characterize the 
tails of $P_{\text{E}}(n)$ and $P_{\text{S}}(\ln n)$ by means of Eqs.~(\ref{PS}) and~(\ref{PEmu}), 
respectively, both scattering functions showed clear non-universal
and parameter-dependent behavior, see e.g.~Figs.~\ref{Fig06a} to~\ref{Fig07b}. 
That is, neither $P_{\text{E}}(n)$ nor $P_{\text{S}}(\ln n)$ can be scaled.

It is important to stress that for the RL-fSM we observe the exponential decay of $P_{\text{S}}(n)$ 
even when the phase space is not ergodic (for sufficiently large iteration times), see 
Figs.~\ref{Fig06a} and~\ref{Fig07a}.
This should be contrasted with other maps characterized by a non-ergodic phase space: 
for example, due to stickiness, $P_{\text{S}}(n)$ is known to develop asymptotic power-law tails in the 
case of mixed-chaotic dynamics~\cite{SODB93,LKDCL12} and, due to Fermi acceleration, the decay 
of $P_{\text{S}}(n)$ as a stretched exponential was reported in~\cite{DL12}.

We hope that our results may motivate further numerical as well as theoretical studies on the
scattering properties of fractional dynamical systems in the context of General Fractional Dynamics
(GFDynamics), recently established by Tarasov in~\cite{T21c}.

\appendix 
\setcounter{figure}{0}

\section{}
\label{appendix}

To avoid the saturation of the main text, here we present the Poincar\'e surfaces of section for the RL-fSM 
with the parameters used to compute the survival probability $P_{\text{S}}(n)$ and the frequency of escape 
$P_{\text{E}}(\ln n)$ of Figs.~\ref{Fig02}-\ref{Fig07}.

\begin{figure}
\centering
\includegraphics[width=0.7\columnwidth]{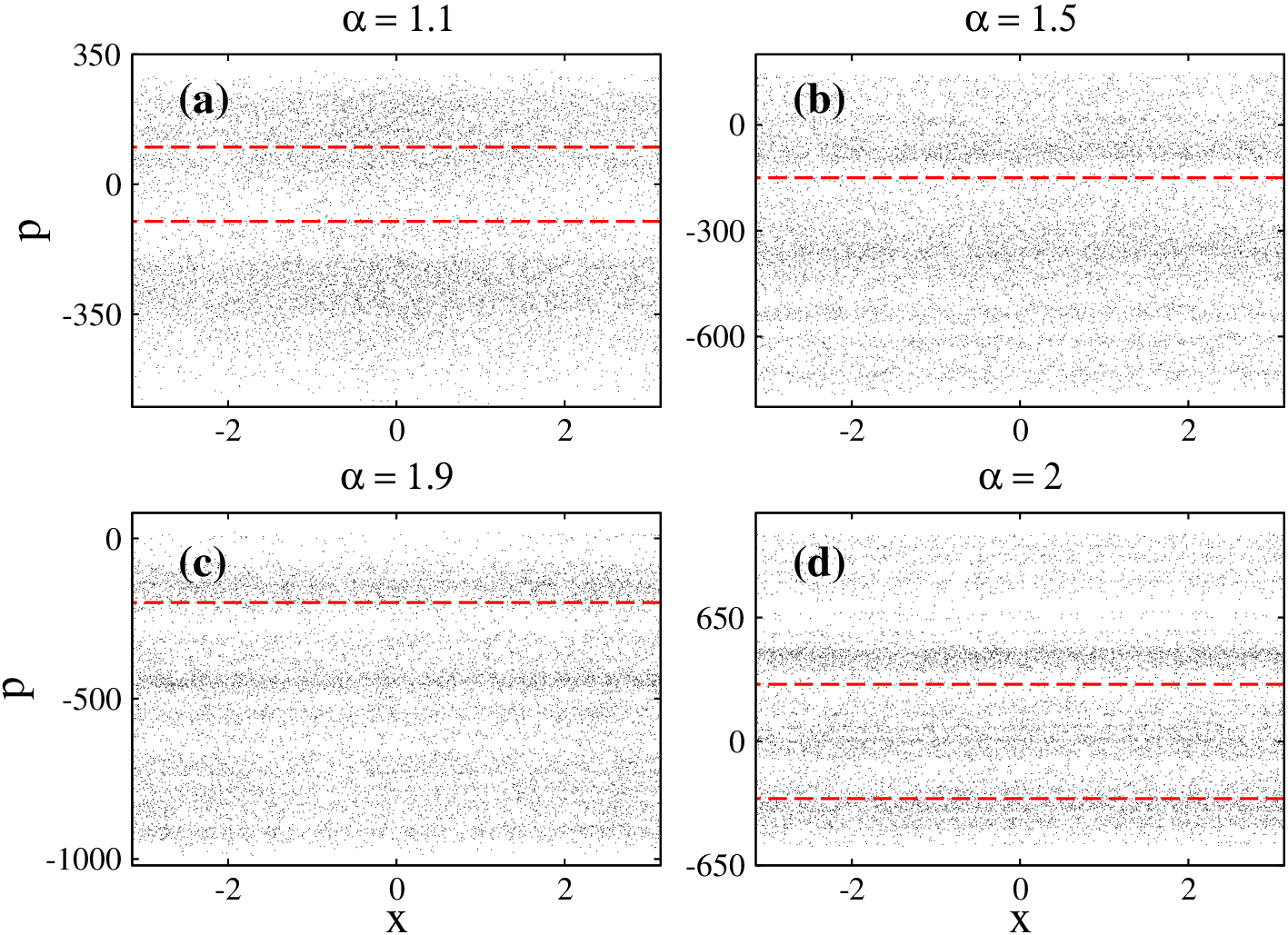}
\caption{Poincar\'e surfaces of section for the RL-fSM with $K = 6.908745$ for the values of $\alpha$
used to compute $P_{\text{S}}(n)$ and $P_{\text{E}}(\ln n)$ of Fig.~\ref{Fig02}. 
A single initial condition with $x_0=0$ and a random $p_0 \in (0,2\pi/100)$ was iterated $10^4$ times.
Holes in the action axis are located at $p=\pm h$ (red dashed lines) with 
(a) $h=100$, (b) $h=150$, (c) $h=200$, and (d) $h=300$.}
\label{FigA01}
\end{figure}
\begin{figure}
\centering
\includegraphics[width=0.7\columnwidth]{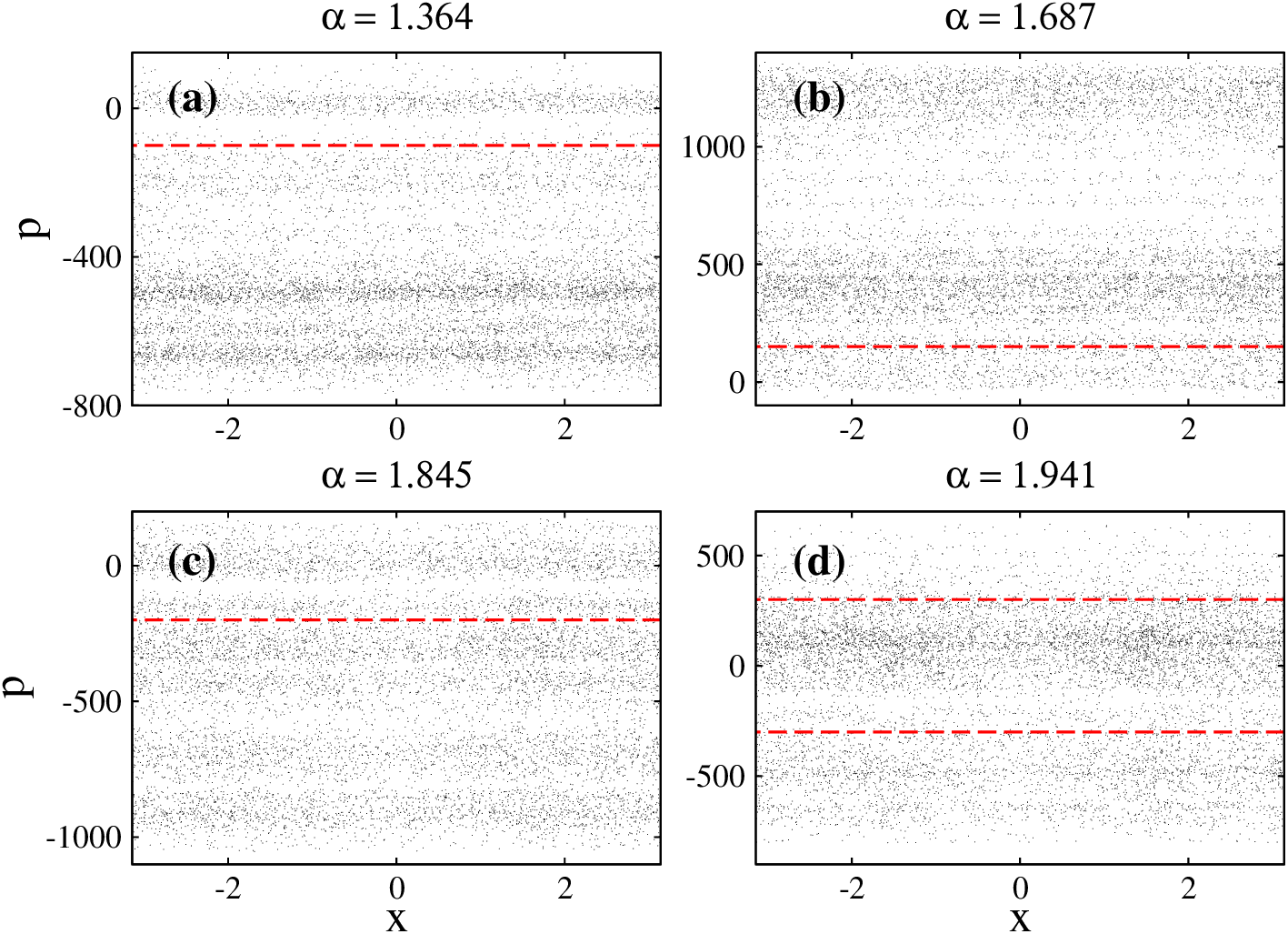}
\caption{Poincar\'e surfaces of section for the RL-fSM with $K = 7.5$ for the values of $\alpha$
used to compute $P_{\text{S}}(n)$ and $P_{\text{E}}(\ln n)$ of Fig.~\ref{Fig03}. 
A single initial condition with $x_0=0$ and a random $p_0 \in (0,2\pi/100)$ was iterated $10^4$ times.
Holes in the action axis are located at $p=\pm h$ (red dashed lines) with 
(a) $h=100$, (b) $h=150$, (c) $h=200$, and (d) $h=300$.} 
\label{FigA02}
\end{figure}
\begin{figure}
    \centering
    \includegraphics[width=0.7\columnwidth]{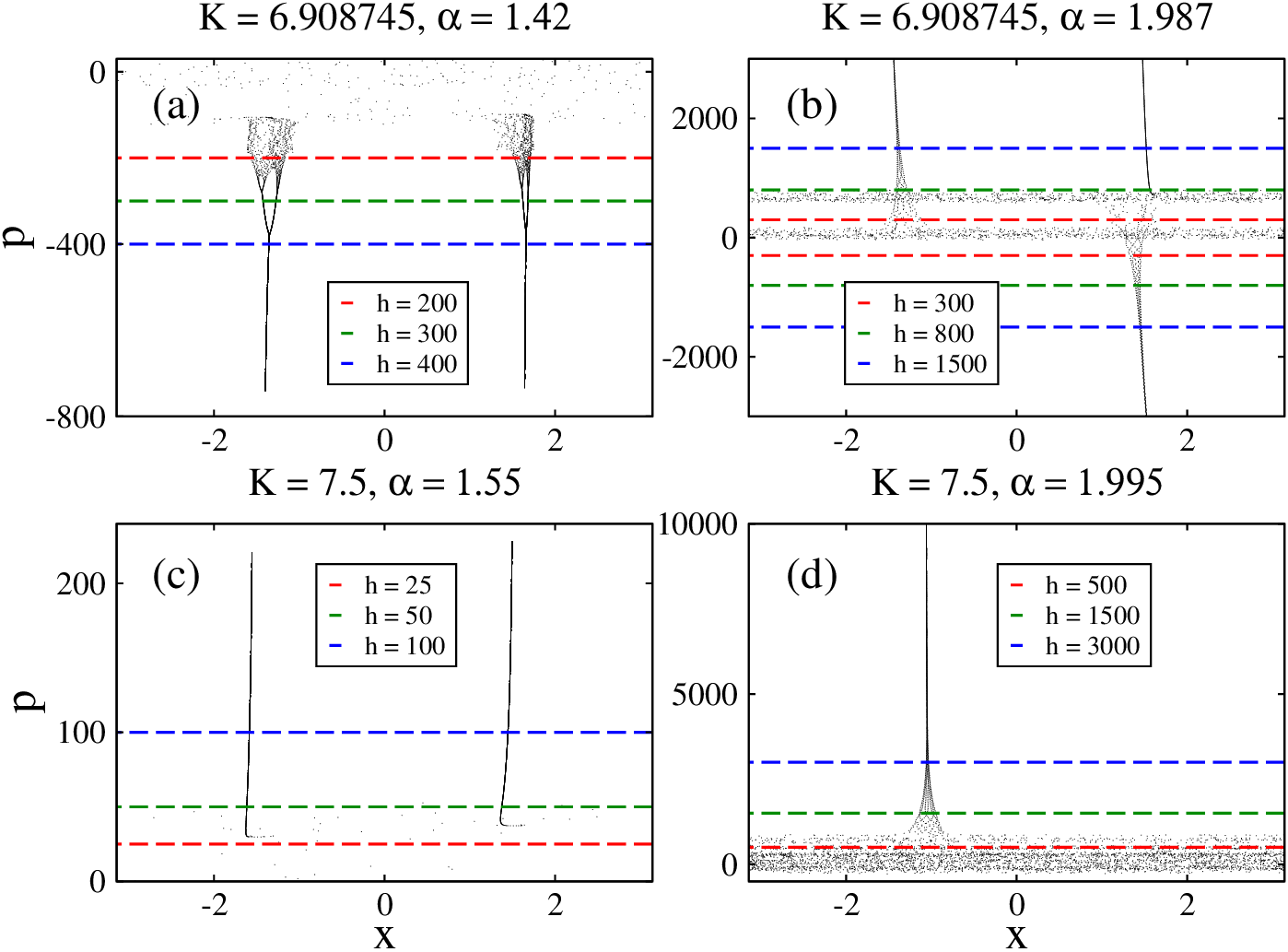}
\caption{Poincar\'e surfaces of section for the RL-fSM with (a,b) $K = 6.908745$ and (c,d) $K = 7.5$
for the values of $\alpha$ used to compute $P_{\text{S}}(n)$ and $P_{\text{E}}(\ln n)$ of Figs.~\ref{Fig06}
and~\ref{Fig07}. 
A single initial condition with $x_0=0$ and a random $p_0 \in (0,2\pi/100)$ was iterated $10^4$ times.
Dashed lines indicate the position of holes in the action axis at $p=\pm h$.}
\label{FigA03}
\end{figure}


\section*{Acknowledgements}

J.A.M.-B. thanks support from CONACyT (Grant No. 286633), CONACyT-Fronteras (Grant No. 425854), 
VIEP-BUAP (Grant No. 100405811-VIEP2022), and Laboratorio Nacional de Superc\'omputo del Sureste 
de M\'exico (Grant No. 202201007C), Mexico.
The research of J.M.S. is supported by a grant from Agencia Estatal de Investigaci\'on 
(PID2019-106433GB-I00/AEI/10.13039/501100011033), Spain.
E.D.L. acknowledges support from CNPq (No. 301318/2019-0) and FAPESP (No. 2019/14038-6), Brazilian 
agencies.


\end{document}